%Paper: cond-mat/9408068
%From: fendley@skat.usc.edu (Paul Fendley)
%Date: Mon, 22 Aug 94 18:36:34 -0700

%
% this file uses harvmac macros, available from cond-mat@babbage.sissa.it
% via "get harvmac.tex" in the subject line
%
\input harvmac
% *********************************************************
\noblackbox
\lref\GZ{S. Ghoshal and A.B. Zamolodchikov, Int. J. Mod.
Phys. A9 (1994) 3841, hep-th/9306002.}
\lref\Zamo{Al.B. Zamolodchikov, Phys. Lett. B253 (1991) 391.}
\lref\YY{C.N. Yang and C.P. Yang, J. Math.  Phys. 10 (1969) 1115;\hfill\break
Al.B. Zamolodchikov, Nucl. Phys. B342 (1991) 695.}
\lref\KM{T.R. Klassen and E. Melzer, Nucl. Phys. B338 (1990) 485.}
\lref\FSW{P. Fendley, H. Saleur and N.P. Warner, ``Exact solution
of a massless scalar field with a relevant boundary interaction'',
to appear in Nucl. Phys. B, hep-th/9406125.}
\lref\Stone{M. Stone and M.P.A. Fisher, ``Luttinger States
at the Edge'', NSF-ITP-94-15, cond-mat/9402040.}
\lref\ZZ{A.B. Zamolodchikov and Al.B. Zamolodchikov, Ann. Phys. 120
(1979) 253.}
\lref\Wen{X.G. Wen, Phys. Rev. B41 (1990) 12838;
Phys. Rev. B43 (1991) 11025; Phys. Rev. B44 (1991) 5708. }
\lref\KF{C.L. Kane and M.P.A. Fisher, Phys. Rev. B46 (1992) 15233.}
\lref\Moon{K. Moon, H. Yi,  C.L. Kane,
S.M. Girvin and M.P.A. Fisher, Phys. Rev. Lett. 71 (1993) 4381.}
\lref\CW{C. De C.Chamon and X.G. Wen, Phys. Rev. Lett.  70 (1993) 2605.}
\lref\FN{A. Furusaki and N. Nagaosa, Phys. Rev. B47 (1993) 3827.}
\lref\ALxray {I. Affleck and A.W.W. Ludwig, ``The Fermi Edge Singularity
and Boundary Condition Changing Operators'', UBCT-94-003, cond-mat/9405057.}
\lref\Lutt{A. Luther and I. Peschel, Phys. Rev. B12 (1975) 3908;
L.P. Kadanoff and A.C. Brown, Ann. Phys. 121 (1979) 318;
F.D.M. Haldane, J. Phys. C14 (1981) 2585.}
\lref\Afflecksusc{I. Affleck, Phys. Rev. Lett. 56 (1986) 2763.}
\lref\FendleyIntr{P. Fendley and K. Intriligator,
Phys. Lett. B 319 (1993) 132, hep-th/9307101.}
\lref\exper{F.P. Milliken, C.P. Umbach and R.A. Webb,
``Evidence for a Luttinger Liquid in the Fractional Quantum
Hall Regime'', IBM preprint 1994.}
\lref\WA{E. Wong and I. Affleck, Nucl. Phys. B417 (1994) 403,
cond-mat/9311040.}
\lref\Guin{F. Guinea, Phys. Rev. B32 (1985) 7518.}
\lref\nonequil{P. Fendley, A.W.W. Ludwig and H. Saleur,
in preparation.}
\lref\FLS{P. Fendley, F. Lesage and H. Saleur,
in preparation.}

%\draft

\def\<{\langle}
\def\>{\rangle}
\Title{\vbox{\baselineskip12pt
\hbox{USC-94-12}\hbox{PUPT-94-1491}\hbox{cond-mat/9408068}}}
{\vbox{\centerline{Exact Conductance through Point Contacts}
\vskip 4pt\centerline{
in the $\nu =1/3$  Fractional Quantum Hall Effect }}}

\centerline{P. Fendley$^1$, A.W.W. Ludwig$^2$ and H. Saleur$^{1*}$ }
\bigskip
\centerline{$^1$ Department of Physics,
University of Southern California}
\centerline{Los Angeles CA 90089-0484}
\medskip
\centerline{$^2$ Joseph Henry Laboratories of Physics,
Princeton University}
\centerline{Princeton NJ 08544}
\vskip.3in

The conductance for  tunneling  through
a point contact between two $\nu =1/3$  quantum Hall edges is
described by
a universal scaling function, which has recently been measured
experimentally. We compute this universal function
exactly, by using the thermodynamic Bethe ansatz and a Boltzmann equation.

\bigskip
\bigskip\bigskip
\noindent $^*$ Packard Fellow
\Date{8/94}

The gapless excitations (edge states) \Wen\
 at the boundary of a fractional Quantum Hall
droplet or bar provide one of the cleanest experimental
realizations \exper\  of a particular  non-Fermi-liquid phase,
often referred to as a Luttinger
liquid. The Luttinger liquid concept \Lutt\
 was originally developed
to describe the excitations about the left and right Fermi points
of an interacting 1D electron gas [1DEG].  However, since
random impurities destroy the non-Fermi-liquid
phase of the 1DEG, it has been difficult to see such
behavior experimentally.
On the other hand, the right- and left-moving edge excitations
of a Quantum Hall bar are identical to those near the right
and left Fermi-points of the 1DEG.  In the
Hall bar, however, these right and left movers are spatially separated,
so that backscattering due to disorder does not affect the
non-Fermi-liquid state, which is therefore stable and experimentally
observable.

One universal signature of the Luttinger state in a
Hall bar of filling fraction $\nu$ is the conductance $G=\nu e^2/h$ \KF.
Another experimental probe of the Luttinger state is the tunneling
conductance through a (local) point contact. The experimental setup
 is a four-terminal
Hall bar geometry (see
\refs{\Moon,\exper}\ for details).
Perpendicular to the $x$-direction of the Hall bar, a
gate voltage $V_g$ induces a tunable  constriction in the $y$-direction,
i.e.\ a point contact.
 By adjusting the gate voltage to a particular value $V_g^*$,
the Hall conductance $G$ (the current in the $x$-direction divided by the
voltage drop in the $y$-direction) is tuned to its maximal (resonance)
value $G=\nu e^2/h$.
Off resonance, $G$ is  predicted to vanish as $T \to 0$ in a universal
way,  described by a universal scaling function of $V_g-V_g^*$
and $T$ \KF. This function, which has recently been measured \exper, is thus
a fingerprint of the Luttinger non-Fermi-liquid state. It is
computed exactly in the following.

 Our method for computing the conductance exactly is based
on the fact that at $\nu=1/3$, the model with the
point-contact interaction is integrable \refs{\GZ,\FSW},
and can thus be studied using exact $S$
matrices and the
thermodynamic Bethe ansatz (TBA) \YY. Integrability does not usually
yield transport properties
at non-zero temperature, because the conventional Kubo
formula requires exact Green functions, which are in general
unknown. Instead, we observe here that integrability defines
a basis of massless
charge-carrying `quasiparticle'
excitations of the non-interacting edges.
These quasiparticles are scattered one-by-one
off the point contact with a momentum-dependent
one-particle scattering matrix $S$
of transmission and reflection amplitudes.
Furthermore, the quasiparticles are characterized by an
exactly-calculable distribution
function. This special behavior is
a consequence of integrability, and it
allows us to derive an exact rate (Boltzmann)
equation for
the conductance in this interacting theory.

\bigskip

We start by describing the left-
and right-moving edge channels at filling
fraction $\nu$
by left and right
moving bosons $\phi_L$ and $\phi_R$. These are defined
on a space $-l < x < l$ (we choose
periodic boundary conditions for convenience).
In the absence of the point-contact interaction,
the dynamics of the bosons are generated
by a Tomonaga Hamiltonian
\eqn \hamil{
H_0={ v_F /\pi\over \nu} \int_{-l}^l dx
\left[j_L^2+j_R^2\right],}
quadratic in the two individually-conserved $U(1)$ currents,
$j_L=-{1\over 4\pi}(\partial_t+\partial_x)\phi_L$ and $j_R={1\over 4\pi}
(\partial_t-\partial_x)\phi_R$. These currents
are the charge densities in the two edges: for example,
$ e Q_L= e\int_{-l}^ldx j_L$ is the total charge
of the left-moving edge.

 Before turning to the backscattering interaction, we review quickly
the calculation of the Hall conductance $G$ for decoupled edges.
We put a voltage drop of $V$ in the $y$-direction, which has
the effect of placing
the charge carriers injected into the
left- and right-moving edges at different chemical
potentials. This is done
by adding a term  $-\Delta Q eV/2 $ to the Hamiltonian, where
$e\Delta Q \equiv e(Q_L - Q_R)$ is the difference between right- and
left-moving charges.
This charge difference is conserved ($\partial_t \Delta Q =0$) when
the edges do not interact, and the resulting
current in the $x$-direction is
\eqn\Idecoupled{
I_0(V)=e v_F
{ \<\Delta Q\>_V\over 2l}.
}
First-order  perturbation theory in $V$ shows
that the linear-response conductance is related to
the charge susceptibility
\eqn\susc{
G_0 =\lim_{V\to 0}{1\over V}I_0(V)=  {e^2  v_F\over 2T}
{ \< \Delta Q \Delta Q \>_{V=0}\over 2 l}
.}
This, in turn, is related \refs{\Afflecksusc, \FendleyIntr}\  to
the chiral
$U(1)$ anomaly $\nu $, defined by
$ \<j_L(x_1) j_L(x_2)\> =$ $\nu/4\pi^2(x_1-x_2)^2$.
 Using these expressions we recover the result of \KF:
$G_0 = \nu e^2 /h$. We use the TBA
to give an alternate derivation below.

We now include a point-contact interaction  coupling
right and left edges at (say) the origin $x=0$.
This is represented in terms
of the bosons $\phi_L,\phi_R$  by  a backscattering contribution to
the Hamiltonian \refs{\Wen,\KF,\Moon,\Stone} :
\eqn \delH {
H_B =\lambda  \{
 e^{i \phi_L(x=0)} e^{-i \phi_R(x=0)}
 +e^{-i \phi_L(x=0)} e^{i \phi_R(x=0)} \}.
}
This can be viewed as the tunneling of Laughlin quasi-particles.
Additional allowed terms are irrelevant when $\nu=1/3$ \KF.
In the presence of this term (i.e.\ for $\lambda \ne 0$)
 there will be an additional
(backscattering)
 contribution $I_B(V)$ to the current,
so that the total current is
 $ I(V) = I_0(V) + I_B(V) $.
We will now derive an exact expression for $I(V)$.
We proceed in four steps. (i) We map the  problem
with Hamiltonian $H_0 + H_B$ into two decoupled
 theories,
one of  which is affected   by the backscattering interaction,
and another one which is not. (ii) We find the
exact quasiparticle spectrum and  exact $S$ matrix,
using the fact that the interacting
theory is integrable.
(iii) We derive an exact (Boltzmann)  equation for the backscattering
current $I_B$. This is possible, despite the non-Gaussian
nature of the interaction, since due to integrability
there is no particle production upon scattering
off the point contact (in an appropriate basis).
(iv) We use the thermodynamic Bethe ansatz to find the exact
distribution functions, allowing us to evaluate the formula
for the conductance explicitly.

\bigskip
Step(i): We see from \delH\  that the interaction involves
 only the combination
$ \phi_L(x=0) - \phi_R(x=0)$.
It is therefore convenient to define  a new basis of {\it even}
and {\it odd} bosons by
\eqn\evenodd{
\phi^e(x+t) \equiv {1 \over \sqrt{2}} [ \phi_L(x,t)+\phi_R(-x,t)],
\qquad
\phi^o(x+t)
\equiv {1 \over \sqrt{2}} [ \phi_L(x,t)-\phi_R(-x,t)]}
($\phi^e$  and $\phi^o$  are  both left-moving, and defined
on $-l < x < l$).
This leads to replacing $j_L,j_R$ with $j^e,
j^o$ in the Hamiltonian \hamil,
where $j^{e/o}(x+t)= (1/\sqrt{2}) [ j_L(x,t) \pm j_R(-x,t)]$
are the charge densities of even and odd bosons.
The even and odd charges  are thus related to
the charges of the original left- and right-moving
edges by
\eqn \evenoddcharge {
\Delta Q = Q_L-Q_R =\sqrt{2} Q^o ,\qquad Q_L+Q_R =  \sqrt{2} Q^e.
}
Two major simplifications arise from this change of
basis. First, the interaction involves
only the odd boson $\phi^o$, while the even boson $\phi^e$
remains free \WA.\foot{There is in fact a residual coupling
between odd and even parts through boundary conditions
\WA\ but it does not affect
the properties we discuss here.}
(This was the crucial step in solving the X-ray edge
problem in a Luttinger liquid \ALxray.)
Additionally, $Q^e$ measures the total charge on both edges and
is conserved even in the presence of the interaction.
Thus the backscattering
current $I_B$ is related to $\Delta Q$  and
can be entirely expressed in terms of the odd boson theory.
Therefore, the even boson will play essentially no role
in the following.

\bigskip
Step (ii): The odd
boson theory is integrable \refs{\GZ,\FSW}.\foot{In the original
 formulations, it rather
appears as a boundary problem. It is well known how
to make an impurity problem into a boundary problem, by
``folding'' the  (odd) system in half.
We shall not do so here and refer to
the treatment of the unfolded problem as in \FSW.}
Integrability means that there is an
infinite number of conserved quantities which commute with
the Hamiltonian, even in the presence of the point
interaction \delH\ \GZ.
The reformulation in terms of the odd boson is vital,
because it is easy to show that
the problem with interaction \delH\ is not solvable in the original
basis of (massless)  Luttinger bosons $\phi_{L,R}$.

There are many bases for
the quasiparticles
of  the odd-boson theory, which are related by not-necessarily-local
mappings. The most obvious
basis for a massless scalar field (the odd boson in our case)
 consists of plane waves, but they are not eigenstates of $H_0+ H_B$
\FSW. A more useful basis is obtained \refs{\GZ,\FSW} by
adding an auxiliary bulk mass term to the Hamiltonian,
which is judiciously chosen so that bulk and  point contact
are both integrable.
This defines a basis of massive bulk excitations
in the odd-boson theory (a massive sine-Gordon model)
which scatter off the point contact without particle production.
By letting the auxiliary bulk mass tend
to zero, a basis of massless particles is obtained.
In this basis the conserved quantities
are $\sum_i p_i^n $, where $p_i$
are the momenta of the individual particles,
and  where $n$ runs over  an infinite  subset  of
the positive integers \GZ.
These conservation laws have important consequences for scattering.
The `quasiparticles' of this basis must scatter
off the point contact {\it without} particle production,
i.e. one-by-one.
Away from the point contact,
the quasiparticles scatter off of each other
with a completely elastic and factorizable scattering matrix $S^{bulk}$.
When $1/\nu$ is an integer, the bulk scattering is diagonal,
so that the only allowed  processes are of the form $ |j(p_1)> \otimes$
$ |k(p_2)>$ $ \to |k(p_2)> \otimes$
$ |j(p_1)>$. Such a process is described by the $S$ matrix element
$S^{bulk}_{jk}(p_1/p_2)$.

The conservation laws in fact allow determination of
the exact quasiparticle spectrum and the $S$ matrix.
This result is already known for
the sine-Gordon model and its massless limit \refs{\ZZ,\GZ}.
The Hamiltonian \hamil\ and \delH\ written
in terms of the odd boson corresponds to the case
$\beta^2=8\pi\nu$ of \FSW. (It is crucial to take
into account the $\sqrt{2}$ in \evenodd.)
At any value of $\nu$, there are a kink ($+$) and an
antikink ($-$). These carry
(odd) charges $ Q^o =
1/\sqrt{2}$ and $-1/\sqrt{2}$,
respectively.
Moreover, for $j-1<1/\nu\le j$, there  are $j-2 $ additional states,
the breathers, which have no charge.
These particles span the Hilbert space
of the left-moving odd boson; we label them
by indices $j,k,..$ running over the kink $(+)$, antikink $(-)$
and breathers $(b)$.

Scattering of a single kink by the point
contact is described by a one-particle $S$ matrix
with elements
$S_{++}(p/T_B)= S_{--}(p/T_B)$ for
 kink $\to$ kink, and antikink $\to$ antikink, as well
as $ S_{+-}(p/T_B)= S_{-+}(p/T_B) $ for
 kink $\to$ antikink, and vice versa.
Here $T_B \propto \lambda^{1/(1-\nu)}$ is
the  crossover scale introduced
by the interaction. These were derived exactly in \GZ:
\eqn \bdrs {\eqalign{
S_{++}(p/T_B)&={(p/T_B)^{(1/\nu)-1}
\over 1+ i{(p/T_B)^{(1/\nu)-1}}}
\exp[i\alpha_{\nu} (p/T_B)] \cr
S_{+-}(p/T_B)&={1\over 1+ i(p/T_B)^{(1/\nu)-1}}\exp[i\alpha_{\nu} (p/T_B)]
.}}
Here  $\exp[i \alpha_{\nu} ]$ is the phase of the expression
given in  Eq.(3.5)
of \FSW\ (where   $\lambda$ there is $(1/\nu) -1$).
These two $S$ matrix elements are interchanged as compared to
\FSW, because we have defined the charges here so
that $S_{++}=1$ at $T_B=0$.
This boundary $S$ matrix is
unitary: $|S_{++}|^2 + |S_{+-}|^2=1$.

\bigskip
Step (iii): We have shown that the bosonic field theory
with Hamiltonian $H_0+H_B$ can be studied in terms
of a particular set of quasiparticles and their scattering.
We now compute an exact equation for the conductance using this basis.

Without the backscattering,
the left and right charges
(or equivalently, the even and odd charges) are conserved individually.
The backscattering allows processes where a charge carrier of the
left-moving edge hops to the right-moving edge or vice versa.
In the original basis, the current $I_B$ is the rate
at which  the charge of the left-moving edge is depleted.
By symmetry, $\partial_t Q_L = - \partial_t Q_R$ in each such hopping
event, so
\eqn \rate {
I_B =  \partial_t  \bigl ( {e\over 2}\Delta Q \bigr ) =
\partial_t  \bigl ( {e\over \sqrt{2}} Q^o \bigr ),
}
and we see that in the even/odd basis, the tunneling
corresponds to the violation of odd charge conservation
at the contact. In the $S$ matrix language
this happens when $S_{+-}\ne 0$, so that
a particle of positive odd charge
(the kink) can scatter into one of negative charge (the antikink) at
the contact. Neutral quasiparticles
cannot transport charge and thus do not
directly contribute to $\partial_t \Delta Q$.

To calculate the conductance, we start with a gas of
quasiparticles with a chemical potential difference
for kinks and antikinks corresponding to the voltage $V$.
A positive voltage means that there are more kinks.
When there are more kinks than antikinks, the backscattering will
turn more kinks to antikinks than it turns antikinks to kinks.
When a kink of momentum $p$
 is scattered into an antikink (the conservation laws
require that it have the same momentum $p$)
this changes $\Delta Q$ by $+2$.
Since kink and antikink quasiparticles scatter off the point contact
one-by-one, we may describe the
 rate at which this charge transport
occurs in terms of two quantities: the probabilities of
finding a kink or antikink of momentum $p$ at the contact, and
the transition probability
$|S_{+-}(p/T_B)|^2$.
We therefore study the density of states $n_V(p)$
and the distribution functions $f_\pm(p,V)$ in the thermodynamic
limit ($l \to \infty$) and in the presence of
an applied voltage $V$.
The number of allowed kink or antikink states per unit length
with momentum  between $p$ and $p+dp$
 is given by $n_V(p) d (v_F p)$,
while the number of states actually occupied by kinks or antikinks
in this momentum range is
$n_V(p) f_+(p,V) d (v_F p) $ and $n_V(p) f_-(p,V) d (v_F p) $,
respectively.
Because at most one of these quasiparticles is allowed per
level \refs{\FSW,\KM}
(a consequence of their being the massless limit of
the sine-Gordon kinks), we have $0\le f_\pm \le 1$.
Thus
\eqn  \DeltaQkink {
{\<  \Delta Q\>_V \over    2l} = \int_0^\infty
d(v_F p)\ n_V(p)\ [f_+(p,V) - f_-(p,V)]
.}
These thermodynamic functions $n_V(p)$ and $f_{\pm}(p,V)$
are different from the free-fermion
functions when  the odd-boson kink theory is  an interacting
Luttinger liquid ($\nu\ne 1/2$),
but they will be derived exactly
from the TBA below.

Having all these definitions in place, it is now easy to
compute the backscattering current from a rate (Boltzmann) equation.
The number of kinks of momentum $p$
which scatter into antikinks per unit time
is given by $|S_{+-}|^2
v_F n_V f_{+}[1-f_{-}]$; the factor $[1-f_{-}]f_{+}$ accounts for the
probabilities
of the initial state being filled and the final state being open.
The rate at which antikinks scatter to kinks is likewise proportional to
$[1-f_{+}]f_{-}$, so the charge changes at a rate  proportional to
$[1-f_{-}]f_{+} - [1-f_{+}]f_{-}= f_+ - f_-$.
Using \rate\ and \DeltaQkink\ we have
\eqn \Boltzmann {
I_B(V)=
-e\int_0^\infty d(v_F p) n_V(p)\ v_F |S_{+-} (p/T_B)|^2
\left[ f_+(p,V) -
 f_-(p,V) \right].}
{}From \Boltzmann \
we obtain the desired backscattering contribution
to the conductance:
\eqn\condi{G_B=
 \lim_{V \to 0} {1\over V} I_B(V)
= -
2e  \int_0^{\infty} {d (v_F p)}\ n_0(p)
 v_F |S_{+-}(p/T_B)|^2 {\partial_{V}} f_+(p,V) \big|_{V =0}.}
The density of states $n_V(p)$
has been evaluated at $V=0$ because $f_+(p,V)  - f_-(p,V)$
is already proportional to $V$. The total conductance is
thus $G = \nu e^2 /h + G_B$.
(Note that \Boltzmann\ may also be
used to compute the conductance $G(V,T)$ at {\it finite} voltage $V$,
i.e.\ far out of equilibrium. We shall report on this shortly \nonequil.)

To check our result, we consider $\nu=1/2$, where
the conductance was previously derived exactly \KF. Here
the odd-boson kinks are simply free fermions \refs{\Guin, \FSW, \ALxray},
so they have the Fermi distribution
function $f_\pm(p,V)=1/[1+\exp(-p/T\pm eV/2T)]$. These fermions do
scatter non-trivially off of the point contact,
with $S$ matrix given  by \bdrs.
The resulting expression for $G_B$ obtained from \condi\  is
identical to the result in sect.\ VIII of \KF.

\bigskip
Step (iv): We proceed by outlining the TBA computation
of the densities of states $n_V(p)$ and the distribution functions
$f_{\pm}(p,V)$ of the massless odd boson theory at
$\nu=1/3$.\foot{One can include the effects of the backscattering
on $n$ and $f$, but this gives only contributions to
$n$ and $f$ vanishing
with the system size $2l$. For example,
free energy $F(T/T_B)$ resulting from \delH\
was calculated in \FSW.
Since the conductance is computed in the limit
$l \to \infty$ these contributions are not relevant here.
} As explained above, these define the occupation numbers
of quasiparticles,
which are scattered without particle production
by the point contact.
The requirement of a periodic boundary condition results in
an equation which relates the densities $n_j$ and $f_j$
of all these quasiparticles
\eqn\ntof{
n_j(p)= {1 \over v_F h} + {1\over 2\pi p} \sum_k
\int_{0}^{\infty} dp'
\Phi_{jk}(p/p') n_k(p') f_k(p'),}
where $\Phi_{jk}(p)=-i (d / d \ln p) \ln S^{bulk}_{jk}(p)$
\YY.
For $\nu =1/3 $ there is only one breather $(b)$ and we have \KM
$$\Phi_{bb}(x)=~2\Phi_{++}(x)=2\Phi_{+-}(x)=
-{4x\over x^2+1}$$
$$\Phi_{b+}(x)=\Phi_{+b}(1/x)=-{4x^3 + 8x\over x^4+4},$$
where the others follow from the symmetry $+\leftrightarrow -$.

One defines
an auxiliary
pseudoenergy variable  $\epsilon_j$
to parametrize $f_j$ via
$f_j\equiv 1/(1+\exp(\epsilon_j-\mu_j/T))$, where
the $\mu_j$ are the chemical potentials: $\mu_+=-\mu_-=eV/2$;
$\mu_b=0$.
By demanding that the free energy at temperature $T$ (expressible
in terms of $f_j$ and $n_j$)  be
minimized, we find  an
equation for $\epsilon_j $ in terms of the (known) bulk
$S$ matrix elements:
\eqn\tba{\epsilon_j(p/T,V/T) = {p\over T} - \sum_k
\int_{0}^{\infty} {dp'\over 2\pi p'}
 \Phi_{jk}(p/p') \ln [1+ e^{\mu_k/T}e^{-\epsilon_k(p'/T,V/T)}].}
Solving this equation for $\epsilon_j$ gives the functions $f_j$.
Even though the breather does not appear in \condi,
it interacts with the kink and antikink and
affects the calculation of $f_\pm$.

We can now evaluate the conductance explicitly.
Using \tba\ and \ntof, it is
easy to see that $\partial_V n_V(p)|_{V=0}=0$, and that
$\partial_{V} f_{\pm} (p,V) |_{V=0}=\pm{ e \over 2T} f [1 - f]$,
where $f(p) \equiv f_{\pm}(p,V=0)$. Thus using \Idecoupled\
and \DeltaQkink\ gives us the TBA relation for
the conductance without a contact:
$$
G_0=\del_V I_0(V)\big|_{V=0}={e^2\over T}
\int_0^\infty {d(v_Fp)}\ n_0(p) v_F[1- f(p)] f(p).
$$
We will verify shortly that this indeed gives the correct answer
$G_0={ \nu e^2/ h}$. Moreover, this and
the unitarity of the $S$ matrix allows us to write the full $G$ as
\eqn \Landauer {
G ={e^2\over T}
 \int_0^\infty d (v_F p) n_0(p) v_F |S_{++}(p/T_B)|^2 [1-f(p)]f(p)
.}
By definition of $\epsilon$, $f(1-f)=-\del_{\epsilon}f$.
The relations \tba\ and \ntof\ imply that $n_j =
{T \over v_F h} \partial_p \epsilon_j (p/T,V=0)$.
Inserting  \bdrs\ into \Landauer, we find our final result for the
conductance at $\nu=1/3$:
\eqn\finalG{\eqalign{G&=-{e^2\over h}\int_0^{\infty}
{dp} {p^4\over p^4+T_B^4}
{\del\epsilon_+(p/T,0)\over \del p}
{\del f(p)\over \del \epsilon_+}\cr
&={e^2\over h}\int_0^\infty
dx\ x^3{4(T_B/T)^4\over (x^4+(T_B/T)^4)^2}
{1\over 1+e^{\epsilon_+(x,0)}},\cr}}
where we integrated by parts and defined the variable $x=p/T$.

As $T_B/T\to \infty$, we have $G\propto(T/T_B)^4$.
Thus it goes to zero
with the correct exponent, as in \KF.
As $T_B/T\to 0 $, we can also evaluate the conductance
explicitly, becoming $[f(0)-f(\infty)]e^2/h$.
The relation \tba\ gives $f(0)=1/3$ and $f(\infty)=0$, and we
indeed  recover $G_0=e^2/3h$.
For $T_B/T$ small, we can
expand $f(x)$ in powers of $x^{4/3}$ based on the
periodicity of the system \tba\ \refs{\Zamo,\FSW}; this means that
$G- G_0 \propto (T_B/T)^{4/3}$, again in agreement with \KF.
In fact, one can compute
all of the perturbative coefficients
using the TBA along with
Jack-polynomial technology \FLS.
To plot the complete function, one solves \tba\
for $\epsilon_+$ and inserts  the result into \finalG.
We have done so numerically to double-precision accuracy.
We compare our results
with the experimental data as well as with the
Monte Carlo results  of  \Moon\ in fig.\ 1.
The agreement between the Monte Carlo
simulation and our exact scaling
curve is excellent.  The exact value of the
universal parameter $K$
(defined so that $G(X)=KX^{-6}$
for $X$ large and $G(X)=(1-X^2)/3$ for $X$ small) is $K=3.3546...$
(where  $X\approx .74313 (T_B/T)^{2/3}$).
(The value $K\approx 2.6$ quoted in \Moon\ seems to have been
slightly underestimated there.) The comparison with
experiments is not completely staightforward, since the
conductance $G_{pc}$ at the resonance peak decreases with
temperature and is well below its resonance value
$ e^2/3h$. This difficulty arises
since two parameters are needed to tune to resonance
in the absence of parity symmetry \KF\ (relevant for the
experiment) whereas only one parameter, $V_g$, has been
varied experimentally. This could be remedied by varying
the magnetic field on the Hall plateau, as well as $V_g$,
in a future experiment.
Nevertheless, the data collapse well onto a single curve,
and the so-obtained experimental scaling curve is in good
agreement with the theoretical exact curve, given the large
scatter of the data in the tail of the resonance curve.
In particular, the data
 clearly show the predicted
$G \propto  T^4/ (V_g-V^*_g)^6$ behavior in the tail.

To conclude, we have computed
experimentally-measured
 transport properties at non-zero
temperature {\it exactly} by using integrability,
 without  the explicit knowledge of Green's functions.
Furthermore, our methods for computing the conductance
can be immediately extended to non-zero voltage,
providing exact transport properties out of equilibrium
\nonequil.
This shows that integrability is a powerful method
of addressing the many strongly-interacting
 problems of modern condensed-matter physics,
 for which reliable answers are notoriously
difficult to come by.
Integrability is
not so exceptional in low dimensions,
and we hope to report on exact results for other
experimentally-important quantities in the near future.

\bigskip

\leftline{\bf Acknowledgements:}

We thank the many participants of the conference SMQFT
(May '94, USC)
 for discussions, and more particularly I. Affleck, M.P.A. Fisher,
F.D.M. Haldane, F.P. Milliken,  N.P. Warner and E. Wong.
This work was supported by the Packard Foundation, the
National Young Investigator program (NSF-PHY-9357207) and
the DOE (DE-FG03-84ER40168).

\listrefs
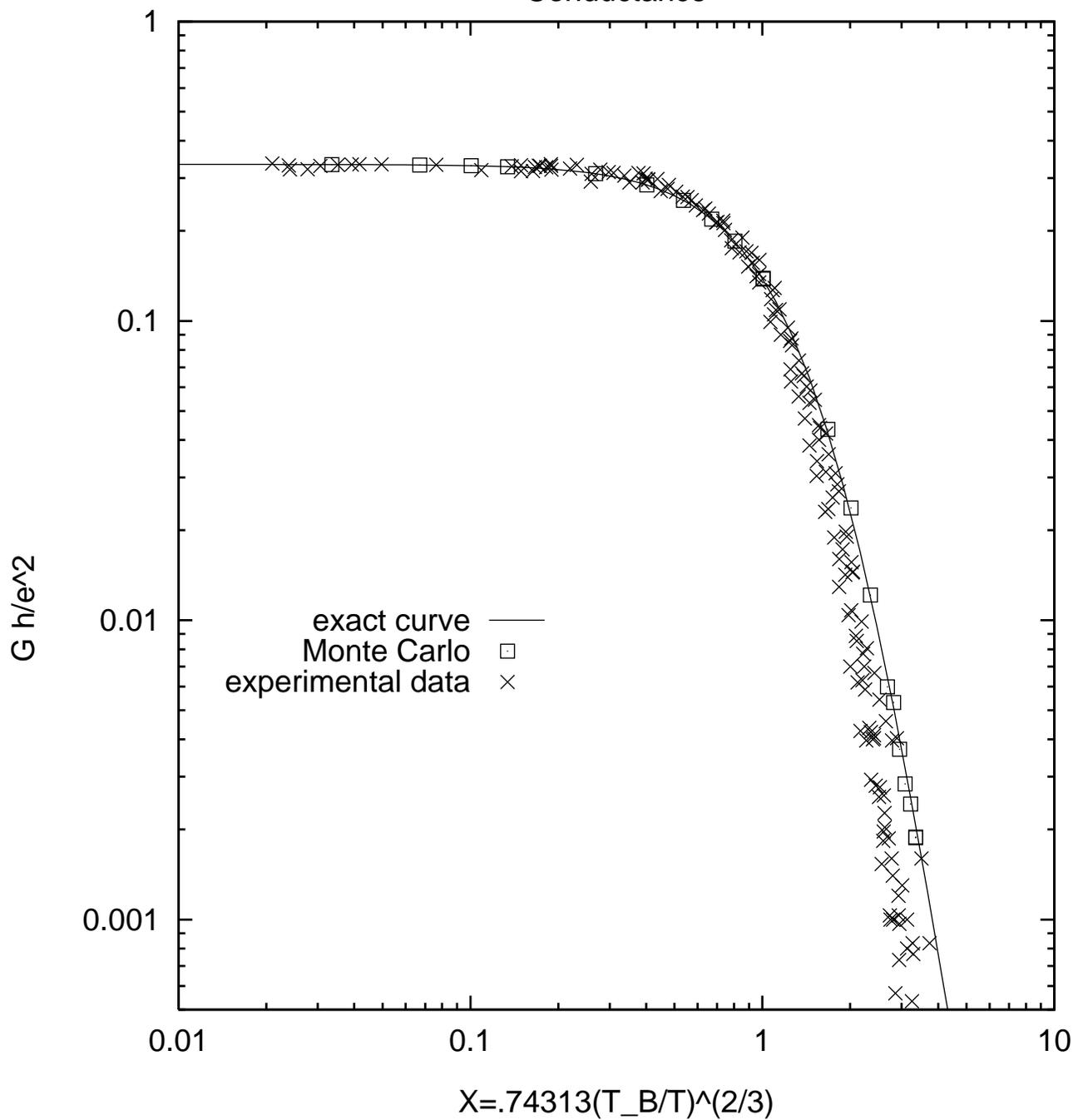
\bye